%% file: main.tex
\pdfoutput=1
\documentclass[12pt]{elsarticle}

\usepackage{caption}
\usepackage{amsmath}
\usepackage{color}
\usepackage[margin=0.75in]{geometry}
\usepackage{fixltx2e}
\usepackage{hyperref}
\usepackage{cleveref}

\makeatletter
\def\ps@pprintTitle{%
   \let\@oddhead\@empty
   \let\@evenhead\@empty
   \let\@oddfoot\@empty
   \let\@evenfoot\@oddfoot
}
\makeatother

\makeatletter
  \long\def\pprintMaketitle{\clearpage
  \iflongmktitle\if@twocolumn\let\columnwidth=\textwidth\fi\fi
  \resetTitleCounters
  \def\baselinestretch{1}%
  \printFirstPageNotes
  \begin{center}%
 \thispagestyle{pprintTitle}%
   \def\baselinestretch{1}%
    \Large\@title\par\vskip18pt
    \normalsize\elsauthors\par\vskip10pt
    \footnotesize\itshape\elsaddress\par\vskip36pt
    \end{center}%
  \gdef\thefootnote{\arabic{footnote}}%
  }
\makeatother

\begin{document}

\begin{frontmatter}

\title{Getting More Out of the Wind:\\ Extending Betz's Law to Multiple Turbines} 

\author[1]{Danny Broberg}
% \cortext[author]{Corresponding author: D.Broberg (dbroberg@berkeley.edu)}
\author[3]{Deep Shah}
\author[4]{Steve Drapcho}
\author[5]{Anna Brockway}

\address[1]{Department of Materials Science and Engineering, University of California, Berkeley, CA 94720, USA}
\address[3]{Department of Chemical Engineering, University of California, Berkeley, CA 94720, USA}
\address[4]{Department of Physics, University of California, Berkeley, CA 94720, USA}
\address[5]{Energy and Resources Group, University of California, Berkeley, CA 94720, USA}

\end{frontmatter}

%Within the software, tools to generate
%and parse density functional calculational results are also included.

\noindent

\input{mainbody}

% \appendix
% \newpage
% \input{appendix}

% \vspace{1in}
% \section*{References}
\bibliographystyle{elsarticle-num}
\bibliography{reference}

\end{document}

%% file: mainbody.tex
\vspace{-.2in}
The Betz law~\cite{betzlaw, dincer2012exergy, munteanu2008optimal} 
places limits on the amount of energy that can be extracted from an incident column of air with a fixed cross sectional area. Some energy is extracted from the column of air, and then the air continues beyond the turbine. This manuscript extends the original Betz Law derivation to a ``multi-turbine'' approach. It begins with a review of the original derivation in Section~\ref{orig} and then generalizes to an $N > 1$ turbine system in Section~\ref{II}. 

\section{Limits on a single turbine (Betz' Law)}\label{orig}

\begin{figure}[h] %this figure will be at the right
    \centering
    \includegraphics[width=0.4\textwidth]{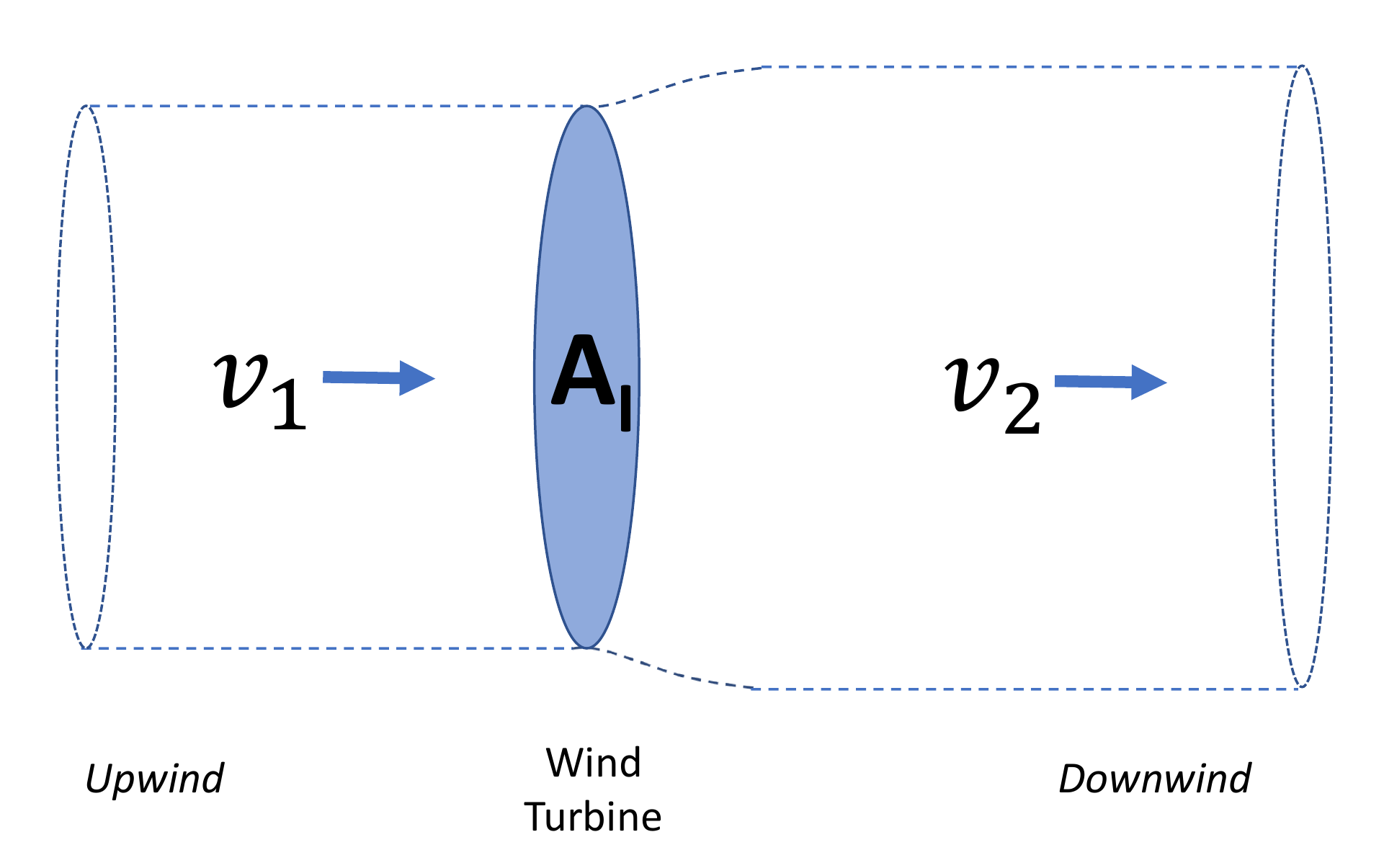}
    \caption{A wind column incident on a single wind turbine}\label{single}
\end{figure}

Consider a column of air, with density $\rho$, moving at velocity $v_1$ incident on a turbine with cross sectional area A, as shown in Figure~\ref{single}.
Since the wind is moving horizontally, the potential energy is more or less constant throughout the wind column and the entirety of the energy is given by kinetic energy, $KE = \frac{1}{2}mv^2$. 

The power at a specific point of the constant velocity column is then given by:
\begin{equation*}
\frac{d}{dt}(KE)=\frac{1}{2}(v^2 \frac{dm}{dt}+2mv \frac{dv}{dt})= \frac{1}{2}v^2 \frac{dm}{dt}
\end{equation*}
where $\frac{dm}{dt}$ is the instantaneous mass flow for a cross section of the column.

If the column of air moving out of the turbine has velocity $v_2$, then power conservation tells us that:
\begin{equation*}
\frac{1}{2} \frac{dm}{dt} v_1^2 = P\textsubscript{turbine} + \frac{1}{2} \frac{dm}{dt} v_2^2
\end{equation*}
By mass flow conservation, the flux of mass must be the same for all cross sections of the column. Therefore the power that is extracted by the turbine (without any additional losses) is 
\begin{equation}\label{pow1}
P\textsubscript{turbine} = \frac{1}{2} \frac{dm}{dt} (v_1^2 - v_2^2)
\end{equation}

Another way we can consider the power into the turbine is from the derivative of the force on the turbine from the incident wind. Consider a test volume with width, $\Delta x$, at the turbine. Then the force performed in that test volume is:
\begin{equation}\label{forceeqn}
\textrm{Force} = \frac{d}{dt}(mv) = (\frac{dm}{dt})_\textsubscript{volume} v + m (\frac{dv}{dt})_\textsubscript{volume} = m \frac{(v_1-v_2)}{\Delta t} = \frac{m}{\Delta t}(v_1 - v_2)
\end{equation}
Where $\Delta t$ is the amount of time it takes for the mass, m, to flow through the volume. The subscripts for the derivatives help to highlight that mass and velocity derivative are taken over the entire test volume.

Next we use the fundamental relation for work (energy) extracted from a (constant) force through this test volume.
\begin{equation*}
E = \int F\cdot x = F \Delta x
\end{equation*}
So that the power is the energy extracted over a time interval, $\Delta t$:
\begin{equation}\label{pow2}
P = \frac{1}{\Delta t} E =  F \frac{\Delta x}{\Delta t} = F v
\Rightarrow P = \frac{dm}{dt}(v_1 - v_2) v
\end{equation}

Equating equations~\ref{pow1} and~\ref{pow2}, allows one to derive an expression for the velocity of the wind speed at the turbine:
\begin{equation*}
\frac{dm}{dt}(v_1-v_2)v = \frac{1}{2}\frac{dm}{dt}(v_1^2-v_2^2) = \frac{1}{2}\frac{dm}{dt}(v_1 + v_2)(v_1 - v_2)
\end{equation*}
\begin{equation}\label{veqn}
\Rightarrow v = \frac{1}{2}(v_1 + v_2) 
\end{equation}

Now note that the column of air flowing through the turbine has a mass flow of: 
\begin{equation}\label{mf}
\frac{dm}{dt}=\frac{d}{dt} (\rho A \Delta x )=\rho A v.
\end{equation}
Where v and A are the instantaneous velocity and cross section. Note that this derivation assumes that air is incompressible and so we do not allow for the density, $\rho$ to change (this would involve additional energy losses that go into the compression/expansion of air).
Combining equations~\ref{pow1},~\ref{veqn} and~\ref{mf} we produce an expression for the turbine power (without any losses):
\begin{equation*}
P_\textsubscript{1 turbine} = \frac{1}{2}\frac{dm}{dt}(v_1^2-v_2^2) = \frac{1}{2}\rho A v (v_1^2-v_2^2) = \frac{1}{4}\rho A (v_1 + v_2)  (v_1^2-v_2^2)
\end{equation*}
 
Dividing this power by the input power, $\frac{1}{2}\rho A v_1^3$, yields the efficiency of a single turbine:
\begin{equation*}
\eta_1 = \frac{P_\textsubscript{1 turbine}}{P_{in}} = \frac{1}{2}\Bigg(1 + \frac{v_2}{v_1}\Bigg)\Bigg(1 - \Big(\frac{v_2}{v_1}\Big)^2\Bigg)
\end{equation*}
Maximizing $\eta$ with respect to the ratio $R_2 = v_2/v_1$:
\begin{equation*}
\frac{d}{dR_2}( \frac{1}{2}(1 + R_2)(1 - R_2^2)) = 0
\end{equation*}
\begin{equation*}
\Rightarrow R_2 = \frac{1}{3}
\end{equation*}

Therefore the ratio $R_2 = v_2/v_1$ which maximizes the efficiency of the turbine is 1/3, yielding an efficiency of $59.2\%$. This is the traditional Betz law limit~\cite{betzlaw}.

\section{ Getting more out of the wind}\label{II}
Now we consider additional turbines behind the original turbine in order to calculate the excess energy which may be extracted from the original incident wind column. This relies on using a larger rotor downwind, to account for the expanded wind column. Since the original Betz law focuses on a fixed swept area, this extension for multiple turbines does not contradict the original formulation by Betz.

\subsection{ Two Turbines}\label{IIA}

\begin{figure}[h] %this figure will be at the right
    \centering
    \includegraphics[width=0.7\textwidth]{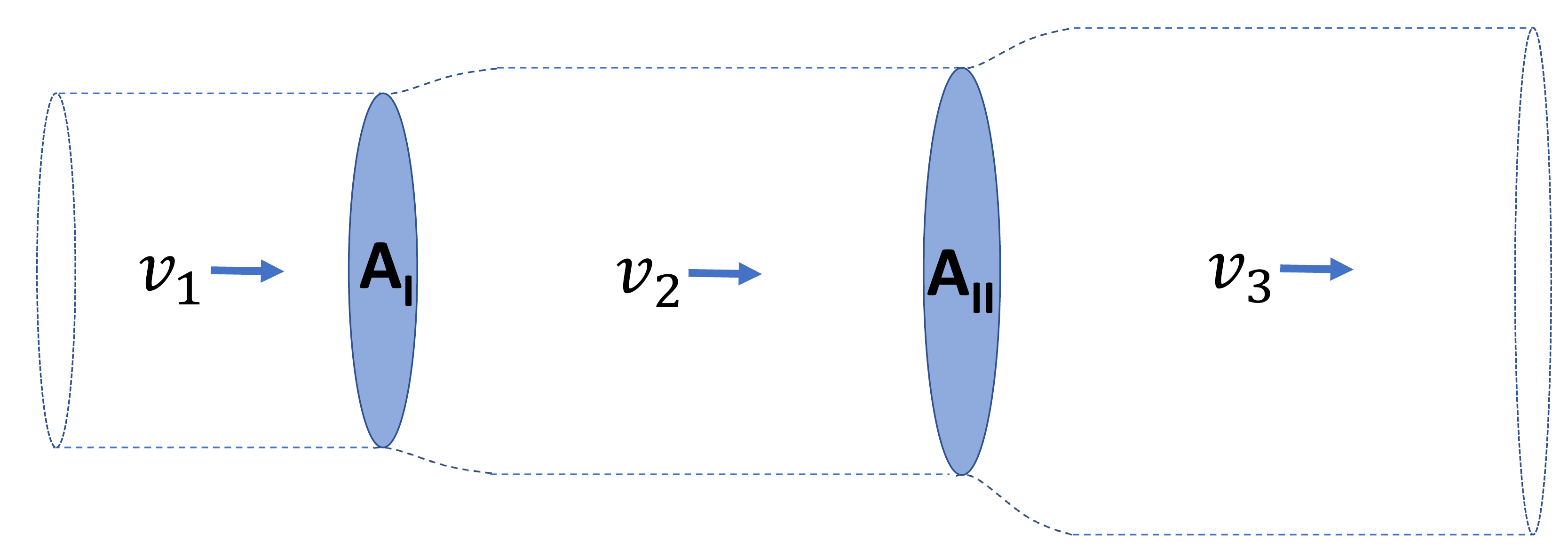}
    \caption{A wind column which is incident on a first turbine expands to a large cross section size, $A_{II}$. It then proceeds through a second turbine with the same cross section.}\label{double}
\end{figure}

As illustrated in Figure~\ref{double}, consider an incident wind velocity $v_1$ into the first turbine with cross sectional area $A_I$, yielding a column of air that expands and has velocity $v_2$. This column of air is then incident on a second turbine with cross sectional area $A_{II}$ (equal to the cross sectional area of the expanded column of air). Once the wind column moves through the second turbine, it has a final velocity $v_3$.

Isolating attention to a single turbine within this multi-turbine system, the derivation for the power extracted without any losses is equivalent to the original Betz law, as outlined in Section~\ref{orig}, assuming that one considers the velocity and cross section from the upwind column of air with the downwind velocity immediately after the turbine. Using the original derivation for each turbine, we therefore have:
\begin{equation*}
\textrm{Lossless Power into turbine 1:  } P_{1} = \frac{1}{4}\rho A_I(v_1 + v_2)(v_1^2 - v_2^2)
\end{equation*}
\begin{equation*}
\textrm{Lossless Power into turbine 2:  } P_{2} = \frac{1}{4}\rho A_{II}(v_2 + v_3)(v_2^2 - v_3^2)
\end{equation*}
Note that besides the velocity indices changing, we are also now changing the indices of the cross section at each turbine. The total power for the combined system is then:
\begin{equation*}
P_\textsubscript{2 turbines} = \frac{1}{4}\rho A_I(v_1 + v_2)(v_1^2 - v_2^2) + \frac{1}{4}\rho A_{II}(v_2 + v_3)(v_2^2 - v_3^2)
\end{equation*}
\begin{equation*}
= \frac{1}{2}\rho A_I v_1^3 \Big( \frac{1}{2}(1 + R_2)(1 - R_2^2) + \frac{1}{2}\frac{A_{II}}{A_I}(R_2 + R_3)(R_2^2 - R_3^2)\Big)
\end{equation*}

Where $R_j = \frac{v_j}{v_1}$ is the $j^{th}$ outgoing velocity relative to the incident velocity. Dividing by the incident power yields an overall efficiency for the two-turbine system of:
\begin{equation}\label{2turb}
\eta_\textsubscript{2 Turbines} = \frac{P_\textsubscript{2 turbines}}{P_{in}} = \frac{1}{2}\Big( (1 + R_2)(1 - R_2^2) + \frac{A_{II}}{A_I}(R_2 + R_3)(R_2^2 - R_3^2) \Big)
\end{equation}

Equation~\ref{2turb} can be further simplified by considering the constraint of constant mass flow throughout the wind column: 
\begin{equation*}
(\frac{dm}{dt})_\textsubscript{1} = \rho A_I v_1 = (\frac{dm}{dt})_\textsubscript{2} = \rho A_{II} v_2
\end{equation*}
\begin{equation}\label{areastoR}
\Rightarrow \frac{A_{II}}{A_I} = \frac{v_1}{v_2} = \frac{1}{R_2}
\end{equation}
Resulting in:
\begin{equation*}
\eta_\textsubscript{2 Turbines} = \frac{P_\textsubscript{2 turbines}}{P_{in}} = \frac{1}{2}\Big( (1 + R_2)(1 - R_2^2) + \frac{1}{R_2}(R_2 + R_3)(R_2^2 - R_3^2) \Big)
\end{equation*}

This is minimized by setting $(d\eta_\textsubscript{2 turbines}/dR_2) = 0$ and solving for $R_2$ in terms of $R_3$, then requiring that $(d\eta_\textsubscript{2 turbines}/dR_3) = 0$. This yields a maximum efficiency of $72.7\%$ with a ratio of $R_2 \approx 0.64$ and $R_3 \approx 0.21$. 

\begin{figure}[h] %this figure will be at the right
    \centering
    \includegraphics[width=0.7\textwidth]{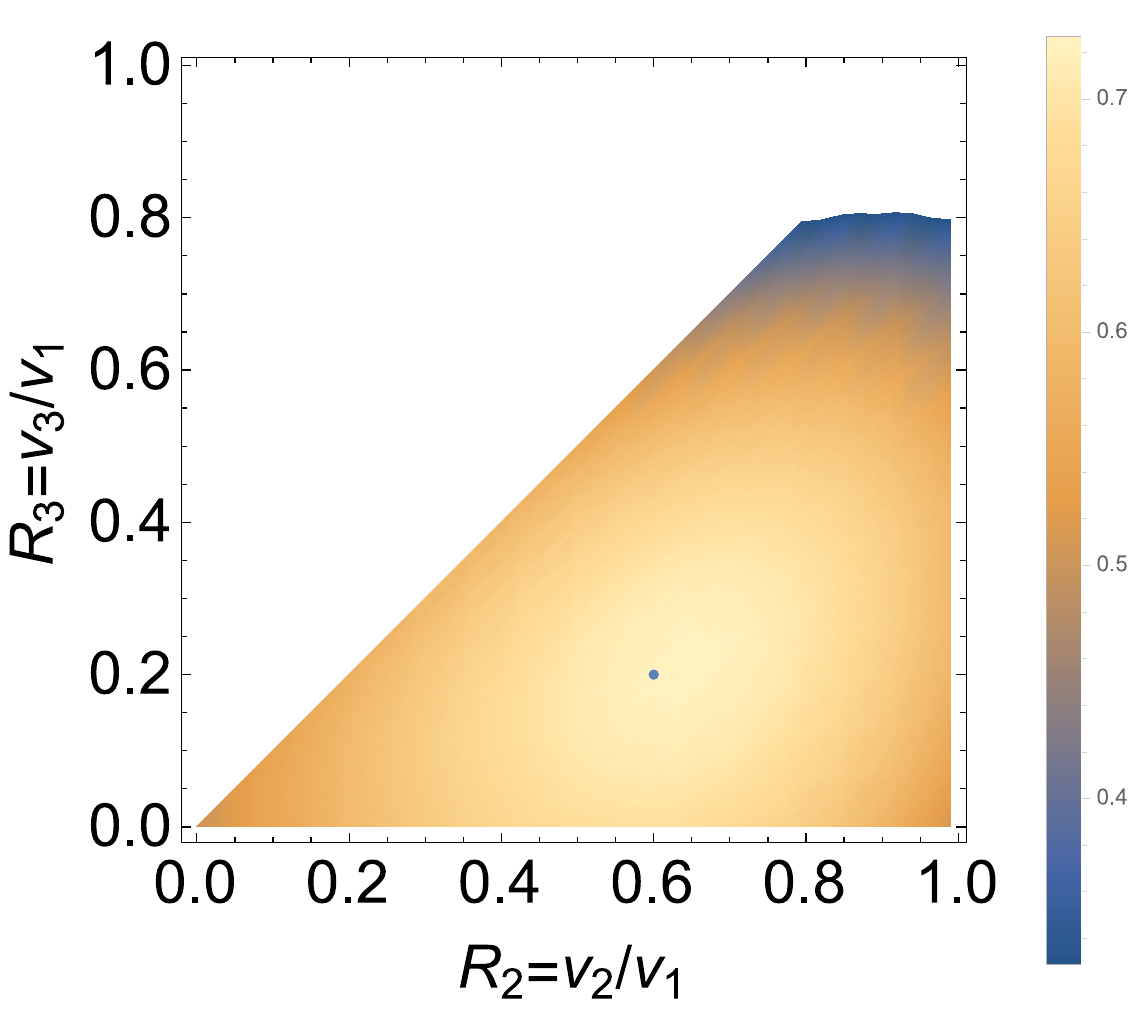}
    \caption{A heat map showing the range of efficiencies for the two turbine system. The maximum efficiency is indicated by a blue dot.}\label{heat}
\end{figure}

Shown in Figure~\ref{heat} is a heat map of the efficiency for a two turbine system for a range of $R_2$ and $R_3$ values. Note that since $v_1 > v_2 > v_3$ is a requirement of positive power extraction, it must be that $R_2 > R_3$.

It is interesting to note that if you isolate your attention to the final wind turbine for the case of maximum efficiency, the ratio of final to incident velocity is  $v_3/v_2 = R_3/R_2 = 1/3$, recovering the Betz law limit for the final turbine, as we would expect.

\subsection{Many ($> 2$) Turbines}\label{IIB}
Generalizing upon the approach of Section~\ref{IIA}, it is straightforward to see that an N-turbine system would have an overall efficiency of:
\begin{equation*}
\eta_\textsubscript{N Turbines} = \frac{1}{2} \sum_{i=1}^N\Big( \frac{1}{R_i}(R_i + R_{i+1})(R_i^2 - R_{i+1}^2) \Big) \textrm{ with } R_j = v_j/v_1
\end{equation*}

The general equation for the $R_j$th derivative term is then (for $2\le j \le N$):
\begin{align}\label{**}
\frac{d\eta_\textsubscript{N Turbines}}{dR_j} &= \frac{1}{2} \frac{d}{dR_j}\Big( \frac{1}{R_{j-1}}(R_{j-1}+R_j)(R_{j-1}^2-R_j^2) +   \frac{1}{R_{j}}(R_j + R_{j+1})(R_j^2 - R_{j+1}^2)
\Big) \nonumber \\
&= \frac{1}{2} \Bigg(\Big(\frac{-3}{R_{j-1}}\Big)R_j^2 + \Big(\Big(\frac{R_{j+1}}{R_j}\Big)^3 +\Big(\frac{R_{j+1}}{R_j}\Big)\Big)R_j + R_{j-1}\Bigg) 
\end{align}

and for $j = N+1$: 
\begin{align}\label{RN1}
\frac{d\eta_\textsubscript{N Turbines}}{dR_{N+1}} 
&= \frac{1}{2} \frac{d}{dR_{N+1}}( \frac{1}{R_{N}}(R_{N}+R_{N+1})(R_{N}^2-R_{N+1}^2)) \nonumber\\
&= \frac{1}{2}\frac{1}{R_N} (R_N^2 - 2R_{N+1}R_N-3 R_{N+1}^2) = 0 \nonumber\\
&\Rightarrow \frac{R_{N+1}}{R_N} = \frac{1}{3} 
\end{align}

Notice that given a ratio of $R_j$ and $R_{j+1}$, call it $U_j = R_{j+1}/R_j$, it is possible to calculate $R_{j-1}$  by setting Equation~\ref{**} to zero and solving for $R_{j-1}$, resulting in:
\begin{align}\label{finalrelation}
\Rightarrow R_{j-1} = \frac{1}{2} R_j (-U_j - U_j^3 + \sqrt{12 + U_j^2 +2U_j^4 +U_j^6}) 
\end{align}

If one begins with the known fact that $R_{N+1}/R_N = 1/3$ (Equation~\ref{RN1}) then they can use equation~\ref{finalrelation} to compute $R_{N-1}$. This process is repeated until $R_1=1$ is reached, at which point a solution for all ratios is reached. This defines an iterative process for solving for the maximum rate of power extraction of an arbitrary N-turbine system.

\begin{figure}[h] %this figure will be at the right
    \centering
    \includegraphics[width=0.8\textwidth]{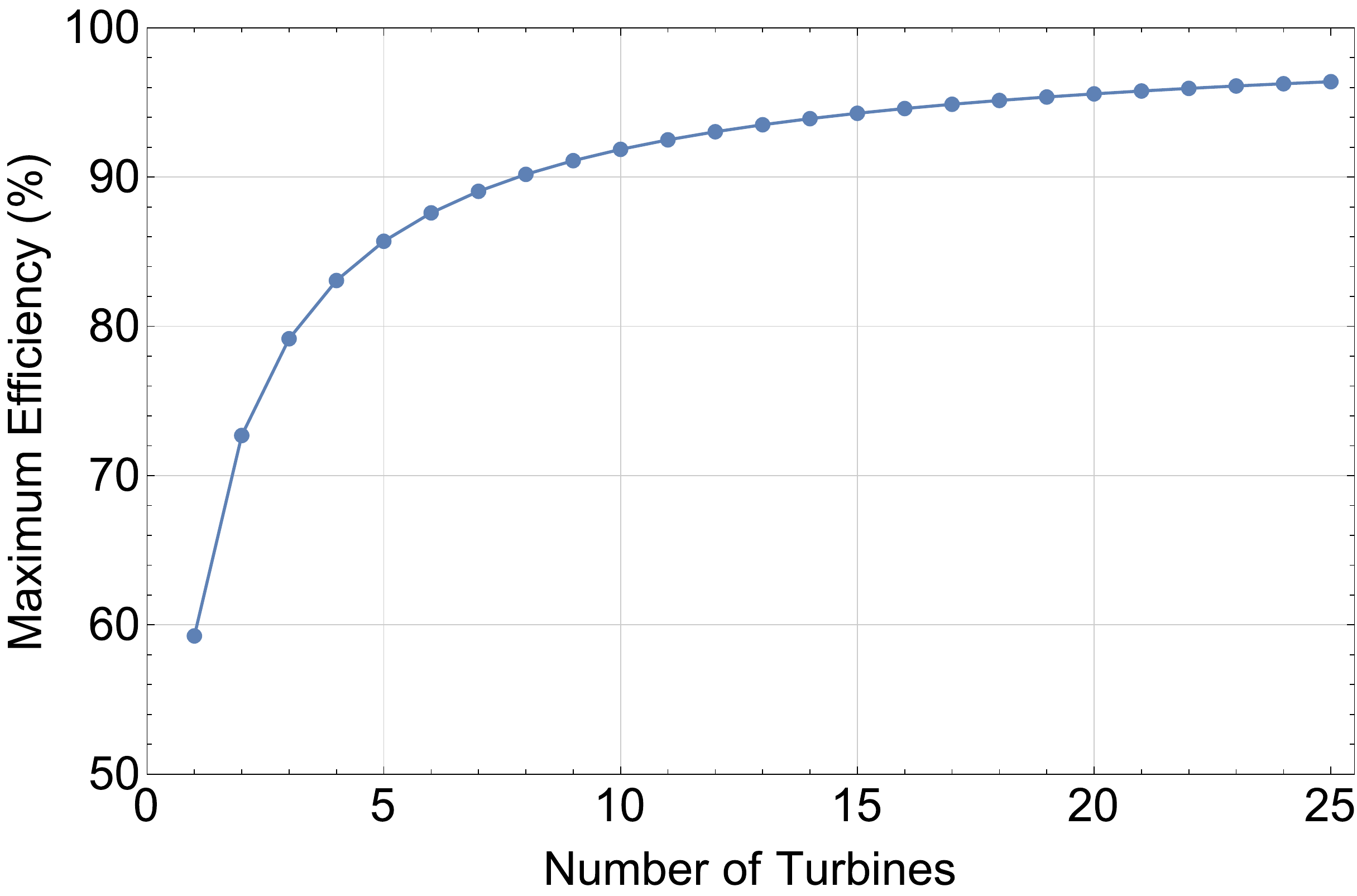}
    \caption{Maximum Efficiencies for up to N=25 turbines.}\label{efficiency}
\end{figure}

\begin{figure}[h] %this figure will be at the right
    \centering
    \includegraphics[width=.99\textwidth]{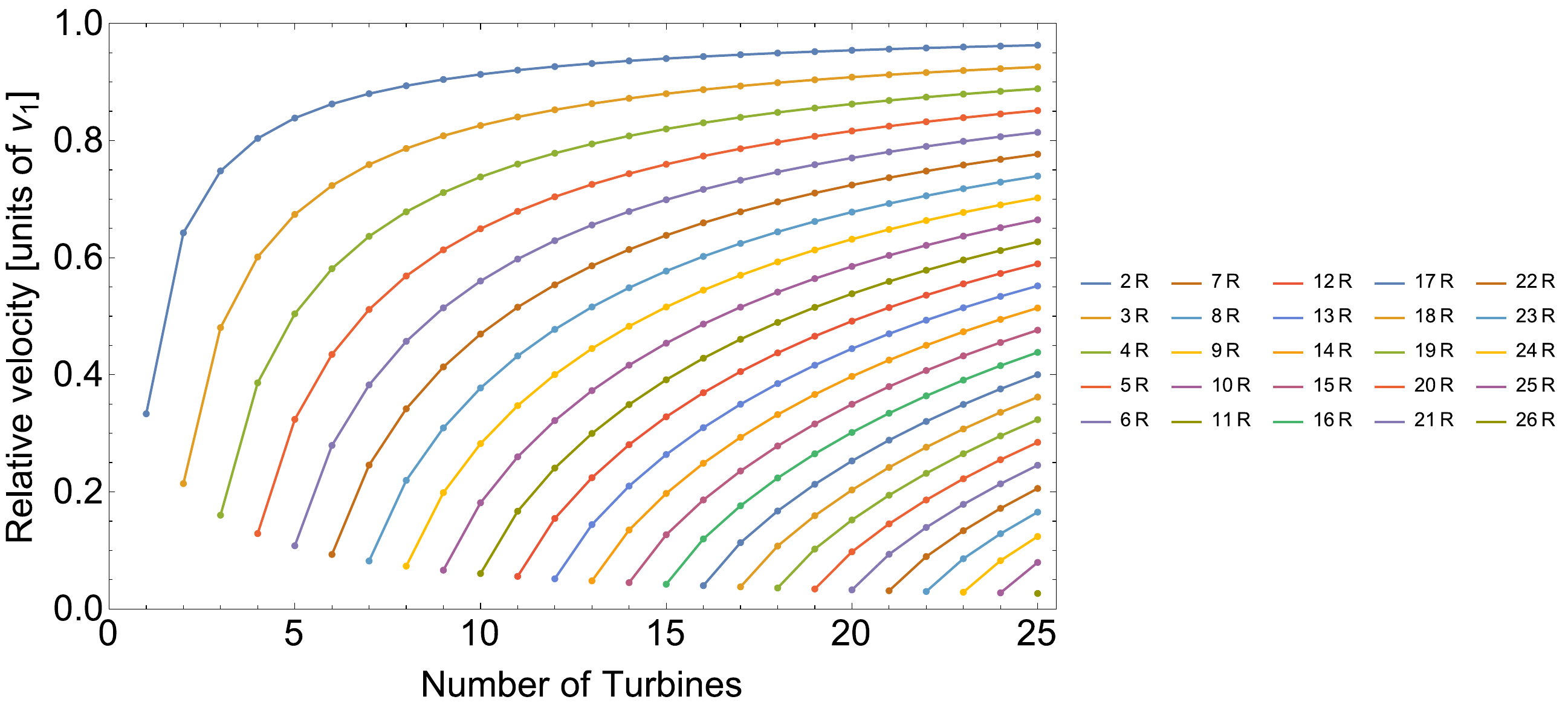}
    \caption{Relative velocities for the optimal velocity ratios for up to N=25 turbines.}\label{relvels}
\end{figure}

This process was calculated for up to 25 turbines and Figure~\ref{efficiency} shows the maximum theoretical rate of power extraction from a column of air approaching $97\%$ with a 25-turbine system. Figure~\ref{relvels} displays the resulting optimal ratios of velocities for each of the 25 turbines. It is interesting to note that the final relative ratios of velocities are monotonically decreasing as the number of turbines increases. This corresponds to a large increase in the cross sectional area of the downwind turbines, since $A_j/A_I = 1/R_j$ (Equation~\ref{areastoR}).  For just the two turbine system, $R_2=0.64$, so that the cross section of the second turbine must be $55\%$ larger than the cross section of the first turbine, while for the 25 turbine system the final turbine must be $12\times$ the cross section of the first turbine. 
% Figure~\ref{relareas} illustrates the cross sectional size increase for the final turbine relative to the initial turbine.

% \vspace{.3in}
\newpage
\section{Conclusion}

This manuscript has illustrated the extension of the original Betz limit for the maximum efficiency of a single wind turbine to a system with multiple turbines. Since the original formulation of Betz's law was based on a fixed swept area, this extension does not negate the original derivation. The authors would like to acknowledge that there is little practical utility for this extended turbine approach, as several unrealistic assumptions are intrinsic to this derivation - such as the requirement for constant momentum of the air column after proceeding through the turbine or the requirement that the density of air remains unchanged throughout the system. Moreover, implementing a multi-turbine system to extract additional power out of one moving column of air is likely infeasible from a cost perspective. Regardless, this extension of the original derivation remains an intriguing thought experiment.

% \begin{figure}[h] %this figure will be at the right
%     \centering
%     \includegraphics[width=0.7\textwidth]{relareas}
%     \caption{Relative cross sectional area of the final turbine, relative to the initial turbine area for up to N=25 turbines.}\label{relareas}
% \end{figure}